\documentclass[prd,showpacs,showkeys,nofootinbib,floatfix,eqsecnum,
               fleqn,preprint,12pt,tightenlines]{revtex4} %for preprint

\usepackage{amsmath,amssymb,revsymb,graphicx,dcolumn}
\newcommand{\version}{v3}  %%true version v2.991
\newcommand{\beq}{\begin{equation}}
\newcommand{\eeq}{\end{equation}}
\newcommand{\beqa}{\begin{eqnarray}}
\newcommand{\eeqa}{\end{eqnarray}}
\newcommand{\bsubeqs}{\begin{subequations}}
\newcommand{\esubeqs}{\end{subequations}}

\begin{document}

\begin{widetext}
\noindent arXiv:1612.04235 \hfill   KA--TP--40--2016 \;(\version)
\newline\vspace*{3mm}
\end{widetext}

\title{More on cold dark matter from $q$-theory\vspace*{1mm}}

\author{F.R. Klinkhamer}
\email{frans.klinkhamer@kit.edu}

\affiliation{Institute for Theoretical Physics,
Karlsruhe Institute of Technology (KIT),\\
76128 Karlsruhe, Germany}

\author{G.E. Volovik}
\email{volovik@ltl.tkk.fi}
\affiliation{\mbox{Low Temperature Laboratory, Department of Applied Physics,}\\
\mbox{Aalto University, PO Box 15100, FI-00076 Aalto, Finland,}\\
and\\
\mbox{Landau Institute for Theoretical Physics,
Russian Academy of Sciences,}\\
\mbox{Kosygina 2, 119334 Moscow, Russia}\vspace*{4mm}}

\begin{abstract}
\vspace*{1mm}\noindent
We consider the rapidly-oscillating part of a $q$-field
in a cosmological context and find that its energy density
behaves in the same way as a cold-dark-matter component,
namely proportional to the inverse cube of the cosmic scale factor.
\newline
\end{abstract}

\keywords{dark matter,\;
dark energy,\;
cosmological constant,\;
cosmology
\vspace*{0mm}
}

\maketitle

%%\newpage%%tmp
\section{Introduction}
\label{sec:Introduction}

In a recent article~\cite{KV2016-q-DM}, we have explored the
idea that the inferred cold-dark-matter component of the
present universe corresponds to the rapidly-oscillating
part of a so-called $q$-field~\cite{KV2008a,%
KV2008b,KV2016-q-brane,KV2016-Lambda-cancellation,%
KV2016-q-ball}.
The constant (spacetime-independent)
$q$-field cancels Planck-scale contributions to the
gravitating vacuum energy density, where the cancellation occurs
without fine tuning.
In this way, there would be a
\emph{combined} solution to the missing-mass problem~\cite{PDG2016}
and the cosmological constant problem~\cite{Weinberg1988},
together with a possible explanation of the nature of
the inferred ``dark-energy'' component of the present
universe~\cite{PDG2016}.

In Ref.~\cite{KV2016-q-DM}, we started from the static equilibrium
configuration of the $q$-field. Here, we turn to a
cosmological context with an evolving universe. The goal is
to establish whether or not the rapidly-oscillating $q$-field
component really behaves as a cold-dark-matter component.

%%\newpage%%tmp
\section{Theory}
\label{sec:Theory}

The results of $q$-theory do not depend on the particular realization
of the variable $q$ which describes the quantum vacuum.
Here, we use the 4-form realization of $q$-theory
with the following action from Ref.~\cite{KV2016-q-DM}:
\begin{subequations}\label{eq:action-qdefinition}
 \begin{eqnarray}
S&=&- \int_{\mathbb{R}^4}
\,d^4x\, \sqrt{-g}\,\left(\frac{R}{16\pi G_{N}} +\epsilon(q)
\right.
%%\nonumber\\[1mm]&&
\left.
+\frac{1}{2}\,C(q)\, g^{\alpha\beta}\, (\nabla_\alpha\, q)\,(\nabla_\beta\, q)
+\mathcal{L}^\text{\,SM}\right),
\label{eq:action}
\\[1mm]
F_{\alpha\beta\gamma\delta} &\equiv&
\nabla_{[\alpha}A_{\beta\gamma\delta]}\,,
\quad
F_{\alpha\beta\gamma\delta}
=q\,\sqrt{-g} \,\epsilon_{\alpha\beta\gamma\delta}\,,
\label{eq:qdefinition}
\end{eqnarray}
\end{subequations}
where $A$ is a 3-form gauge field with a
corresponding 4-form field strength $F \propto q$
(see Refs.~\cite{KV2008a,KV2008b} and further references therein),
$\epsilon(q)$ is a generic even function of $q$,  %%AAA
and $\mathcal{L}^\text{\,SM}$ is the Lagrange density of the fields
of the standard model (SM) of elementary particle physics.
The possible role a 4-form field for the
solution of the cosmological constant problem
has been emphasized by Hawking~\cite{Hawking1984} among others.
Throughout, we use natural units with $c=\hbar=1$
and take the metric signature $(-+++)$.

In order to simplify the analysis as much as possible,
we use these \textit{Ans\"{a}tze}:
\bsubeqs\label{eq:assumptions}
\beqa
\label{eq:assumption-C}
C(q) &=& \text{constant} =(q_{0})^{-1} >0\,,
\\[2mm]
\label{eq:assumption-q0}
q_{0}&=&(E_P)^2
\equiv (G_{N})^{-1}
\approx \left(1.22 \times 10^{19}\,\text{GeV}\right)^2\,,
\\[2mm]
\label{eq:assumption-epsilon}
\epsilon(q)&=&
\frac12\,(q_{0})^2 \,
\left[ \frac13\,\left(\frac{q}{q_{0}}\right)^4
- \left(\frac{q}{q_{0}}\right)^2 \,\right]\,,
\\[2mm]
\label{eq:assumption-mu0}
\mu_{0} &=& - \frac13\,q_{0}\,,
\eeqa
\esubeqs
where $q_0$ is the equilibrium value of the $q$-field
in the Minkowski vacuum
and $\mu_{0}$ is the corresponding equilibrium value of the
integration constant $\mu$ of the generalized Maxwell equation.
This constant $\mu$ also enters the definition of
the gravitating vacuum energy density,
\beq
\label{eq:rho-V}
\rho_{V}(q)\equiv \epsilon(q) - \mu\,q \,,
\eeq
which suggests the interpretation of
$\mu$ as a ``chemical potential''
(see Refs.~\cite{KV2008a,KV2008b} for further discussion).

The reduced Maxwell equation~(4) in Ref.~\cite{KV2016-q-DM},
with integration constant $\mu$, is now precisely
a Klein--Gordon equation,
\begin{equation}
\label{eq:Klein-Gordon-eq-q}
\Box\,q = q_{0}\, \frac{d\rho_{V}(q)}{dq} \,,
\end{equation}
with $\rho_{V}(q)$ from \eqref{eq:rho-V}.
As mentioned already in Ref.~\cite{KV2016-q-ball},
Eq.~\eqref{eq:Klein-Gordon-eq-q} describes, in the equilibrium vacuum,
the spectrum of a massive particle with mass-square $M^2 = q_0$.
Note that the right-hand-side of \eqref{eq:Klein-Gordon-eq-q}
would contain further nonlinear $q$ terms
if $C(q)$ were nonconstant.   %%AAA

The standard Einstein equation,
\begin{equation}
\label{eq:Einstein-eq}
R_{\alpha\beta} - \frac12\, g_{\alpha\beta}\,R = - 8\pi G_{N}
\left( T_{\alpha\beta}^{\,(q)} + T_{\alpha\beta}^\text{\,(SM)} \right) \,,
\end{equation}
has, with the above assumptions,
the following $q$-field energy-momentum tensor~\cite{KV2016-q-DM}:
\begin{eqnarray}
 \label{eq:energy-momentum-tensor-q-followup}
  T_{\alpha\beta}^{\,(q)}
  &=&-\, g_{\alpha\beta} \left[ \rho_{V}(q) 
  + \frac12 (q_{0})^{-1} \, \nabla_\alpha \, q \, \nabla^\alpha q  \right]
%% \nonumber\\[1mm]&&
  + (q_{0})^{-1}\, \nabla_\alpha \, q \, \nabla_\beta \, q \,.
\end{eqnarray}

%%\newpage%%tmp
\section{Cosmological model}
\label{sec:Cosmological-model}

Consider the spatially-flat ($k=0$) Robertson--Walker (RW) metric
for standard comoving coordinates.
The $q$-field is taken to be homogeneous, so that $q=q(t)$.
In the present article, we omit the matter described by the SM fields,
but their relativistic and
nonrelativistic components can easily be added to the
dynamic equations below.

In fact, the reduced Maxwell equation~\eqref{eq:Klein-Gordon-eq-q} and
the standard Einstein equation~\eqref{eq:Einstein-eq} with the $q$-field
energy-momentum tensor \eqref{eq:energy-momentum-tensor-q-followup}
take the following form in a spatially-flat RW universe:%
\bsubeqs\label{eq:dimensional-ODEs}
\beqa
\ddot{q} + 3\, (\dot{a}/a)\,\dot{q} &=&-q_{0}\,\frac{d\rho_{V}(q)}{dq}\,,
\\[2mm]
\ddot{a}/a&=&-\frac{8\pi G_N}{3}
\left[\phantom{\frac12}\hspace*{-1mm}
(q_{0})^{-1}\,(\partial_t q)^2 - (q_{0})^{-1}\,\rho_{V}(q) \right]\,,
\\[2mm]
(\dot{a}/a)^2&=&\frac{8\pi G_N}{3}
\left[\frac12\,(q_{0})^{-1}\,(\partial_t q)^2 + (q_{0})^{-1}\,\rho_{V}(q) \right]\,,
\eeqa
\esubeqs
where the dot stands for differentiation with respect to
the cosmic time $t$ and $a(t)$ is the cosmic scale factor.

Now introduce the dimensionless $q$-field perturbation $\xi$ by
\beq
\label{eq:xi}
q(t)/q_{0} = 1 +\xi(t)
\eeq
and define Planck-scale dimensionless units with
\beq
\label{eq:Planck-units}
q_{0}=(E_P)^2=1\,.
\eeq
For example, the dimensionless cosmic time is given by
$\tau \equiv E_P \, t$ and a generic  dimensionless energy density
by $r_{X} \equiv (E_P)^{-4}\,\rho_{X}$.
In these units,
the dimensionless mass of the $q$-field perturbation is $m=1$,
which is also the oscillation frequency of the $\xi(\tau)$ field
in a RW universe (the dimensionless period being $2\pi$).

With $\rho_{V}(q)$ from \eqref{eq:rho-V}
for general chemical potential $\mu$
and the $\epsilon(q)$ \textit{Ansatz} from \eqref{eq:assumption-epsilon},
the dimensionless ordinary differential equations (ODEs) and
the dimensionless vacuum energy density $r_{V}$ are
\bsubeqs\label{eq:dimensionless-ODEs-rV-u0}
\beqa
\label{eq:dimensionless-ODEs-ddot-xi}
\ddot{\xi}+3\,\left(\frac{\dot{a}}{a}\right)\,\dot{\xi}&=&
-\frac{d \,r_{V}}{d\xi}\,\,,
\\[2mm]
\label{eq:dimensionless-ODEs-ddot-a}
\frac{\ddot{a}}{a}&=&-\frac{8\pi}{3}\,
\left[\phantom{\frac12}\hspace*{-1mm}
\dot{\xi}^2 - r_{V}\right]\,,
\\[2mm]
\label{eq:dimensionless-ODEs-Friedmann}
\left(\frac{\dot{a}}{a}\right)^2&=&\frac{8\pi}{3}\,
\left[\frac12\,\dot{\xi}^2 + r_{V} \right]\,,
\\[2mm]
\label{eq:dimensionless-rV}
r_{V}(\xi)
&=&
\frac12\,\xi^2+\frac23\,\xi^3+\frac16\,\xi^4-(u-u_{0})\,(1 +\xi)\,,
\\[2mm]
\label{eq:dimensionless-ODEs-u0}
u_{0} &=& -\frac13\,,
\eeqa
\esubeqs
where the dot now stands for differentiation with respect to
the dimensionless cosmic time $\tau$. The equilibrium value
of the dimensionless chemical potential $u$ is denoted by $u_{0}$.

%%\newpage%%tmp
\section{Analytic solutions}
\label{sec:Analytic-solutions}

The ODEs \eqref{eq:dimensionless-ODEs-rV-u0}
%\eqref{eq:dimensionless-ODEs-ddot-xi},
%\eqref{eq:dimensionless-ODEs-ddot-a}, and \eqref{eq:dimensionless-ODEs-Friedmann}
have special solutions given by certain constant $\xi(\tau)$ functions.
These solutions correspond to Minkowski spacetime for $u=u_0$
and to de-Sitter spacetime for $u<u_0$.
Specifically the solutions are given by
\bsubeqs\label{eq:dimensionless-ODEs-const-sol}
\beqa
\label{eq:dimensionless-ODEs-const-sol-xi}
\xi(\tau)&=&\xi_{\text{const},\,n}\,\,,
\\[2mm]
\label{eq:dimensionless-ODEs-const-sol-h2}
h^2 &\equiv& \left(\frac{\dot{a}}{a}\right)^2=\frac{8\pi}{3}\;
r_{V}\big(\xi_{\text{const},\,n}\big) \,,
\eeqa
with the constant value $\xi_{\text{const},\,n}$
being a solution of the following equation:
\beq
\label{eq:dimensionless-ODEs-const-sol-xi-const-n}
\left.\frac{d \,r_{V}}{d\xi}\,\right|_{\xi=\xi_{\text{const},\,n}}=0\,,
\eeq
\esubeqs
where the discrete index $n$ labels different solutions.
For the particular $r_V$  \textit{Ansatz} \eqref{eq:dimensionless-rV},
Eq.~\eqref{eq:dimensionless-ODEs-const-sol-xi-const-n} is a cubic equation.

The initial boundary conditions
leading to the special solution \eqref{eq:dimensionless-ODEs-const-sol}
include $\dot{\xi}(1)=0$.
For generic initial boundary conditions, the solution can be written
as follows:
\beq
\label{eq:dimensionless-ODEs-generic-sol-xi}
\xi(\tau)=\xi_{\text{const},\,\overline{n}} + \xi_\text{oscill}(\tau) \,,
\eeq
for a particular index $\overline{n}$ so that
$\xi_\text{oscill}(\tau) \to 0$ for $\tau\to\infty$.
Generic solutions \eqref{eq:dimensionless-ODEs-generic-sol-xi}
can be obtained numerically,
as will be shown in Sec.~\ref{sec:Numeric-solutions}.

For later use, we already define
the dimensionless $q$-dark-matter energy density and pressure,
\bsubeqs\label{eq:rDM-pDM}
\beqa
\label{eq:rDM}
r_\text{$q$-DM} &\equiv& \frac12\,\dot{\xi}^2
+\left[ r_{V}(\xi) - r_{V}(\xi_{\text{const},\,\overline{n}})  \right]\,,
\\[2mm]
\label{eq:pDM}
p_\text{$q$-DM} &\equiv& \frac12\,\dot{\xi}^2
- \left[ r_{V}(\xi) - r_{V}(\xi_{\text{const},\,\overline{n}})  \right]\,,
\eeqa
\esubeqs
with $\xi_{\text{const},\,\overline{n}}$ corresponding to
the asymptotic value of the solution $\xi(\tau)$.
Note that the constant part of $r_{V}$
corresponds to an effective cosmological constant,
\beq
\label{eq:overline-r-V-CC}
r_{V-\text{CC}}=-p_{V-\text{CC}}
\equiv r_{V}(\xi_{\text{const},\,\overline{n}})\,.
\eeq
The square bracket on the right-hand-side
of the Friedmann equation~\eqref{eq:dimensionless-ODEs-Friedmann}
then contains precisely the combination
$r_\text{$q$-DM}+r_{V-\text{CC}}$.

%%\newpage%%tmp
\section{Numeric solutions}
\label{sec:Numeric-solutions}

The numeric solutions are obtained from the
two second-order ODEs \eqref{eq:dimensionless-ODEs-ddot-xi}
and \eqref{eq:dimensionless-ODEs-ddot-a},
with initial boundary conditions obeying
\eqref{eq:dimensionless-ODEs-Friedmann}.

\begin{figure}[t]
\vspace*{-0cm}
\begin{center}   %%FIG1-v082.eps --> FIG1-v1.eps=FIG1-v2.eps
\includegraphics[width=0.90\textwidth]{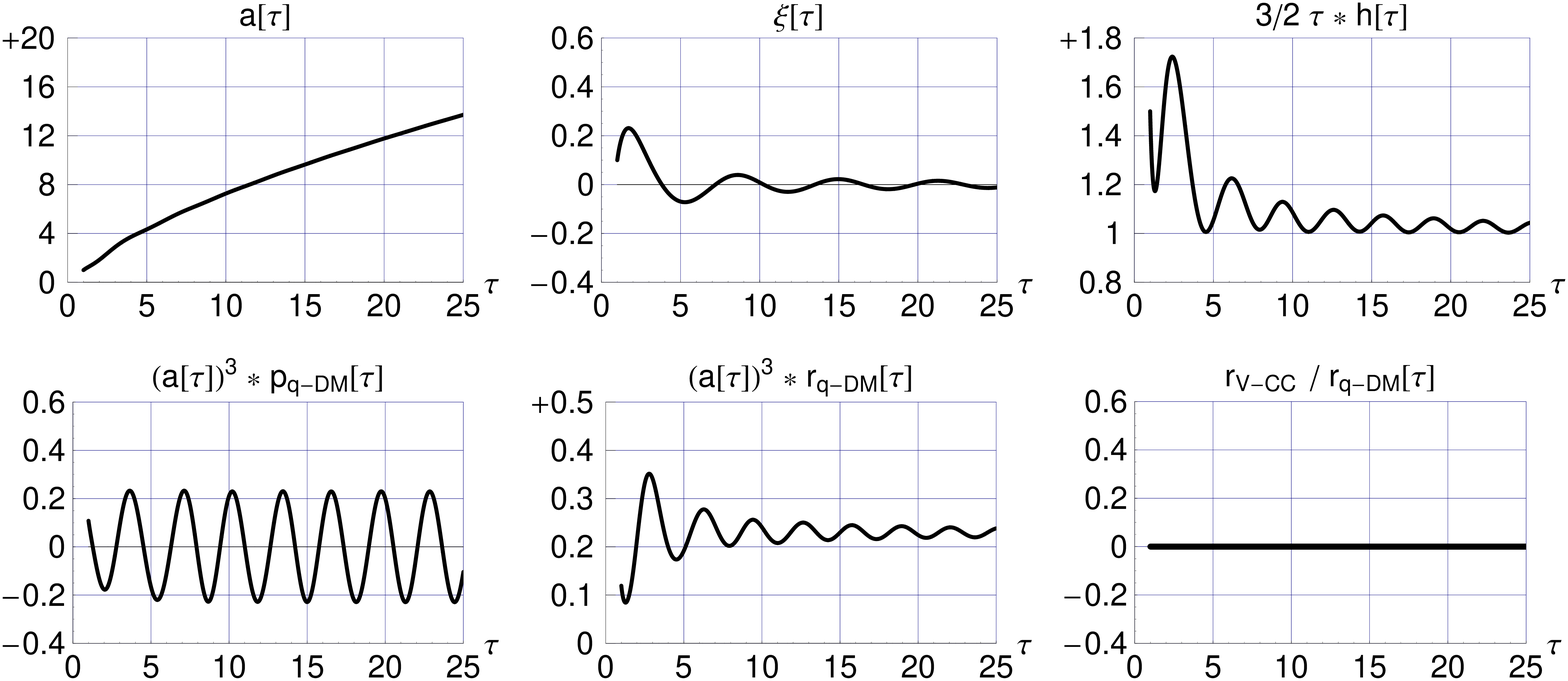}
\end{center}
\vspace*{-5mm}
\caption{Numeric solution of the ODEs \eqref{eq:dimensionless-ODEs-rV-u0}
with $u=u_{0}=-1/3$. The boundary conditions at $\tau=1$ are
$\{a(1),\, \dot{a}(1),\, \xi(1) \}=\{1,\,  1,\, 1/10\}$
with $\dot{\xi}(1)=0.476829$ from \eqref{eq:dimensionless-ODEs-Friedmann}.
The corresponding effective cosmological constant
from \eqref{eq:overline-r-V-CC} vanishes,
with $r_{V-\text{CC}}=-p_{V-\text{CC}}=0$ for $\xi_{\text{const},\,1}=0$.
Shown are the two basic functions $a(\tau)$ and $\xi(\tau)$,
together with the
dimensionless Hubble parameter $h \equiv \dot{a}/a$ and the
dimensionless $q$-dark-matter energy density and pressure from
\eqref{eq:rDM-pDM}.
}
\label{fig:solution-equil-mu}
\vspace*{1cm}
\begin{center}  %%FIG2-v190.eps --> FIG2-v2.eps
\includegraphics[width=0.90\textwidth]{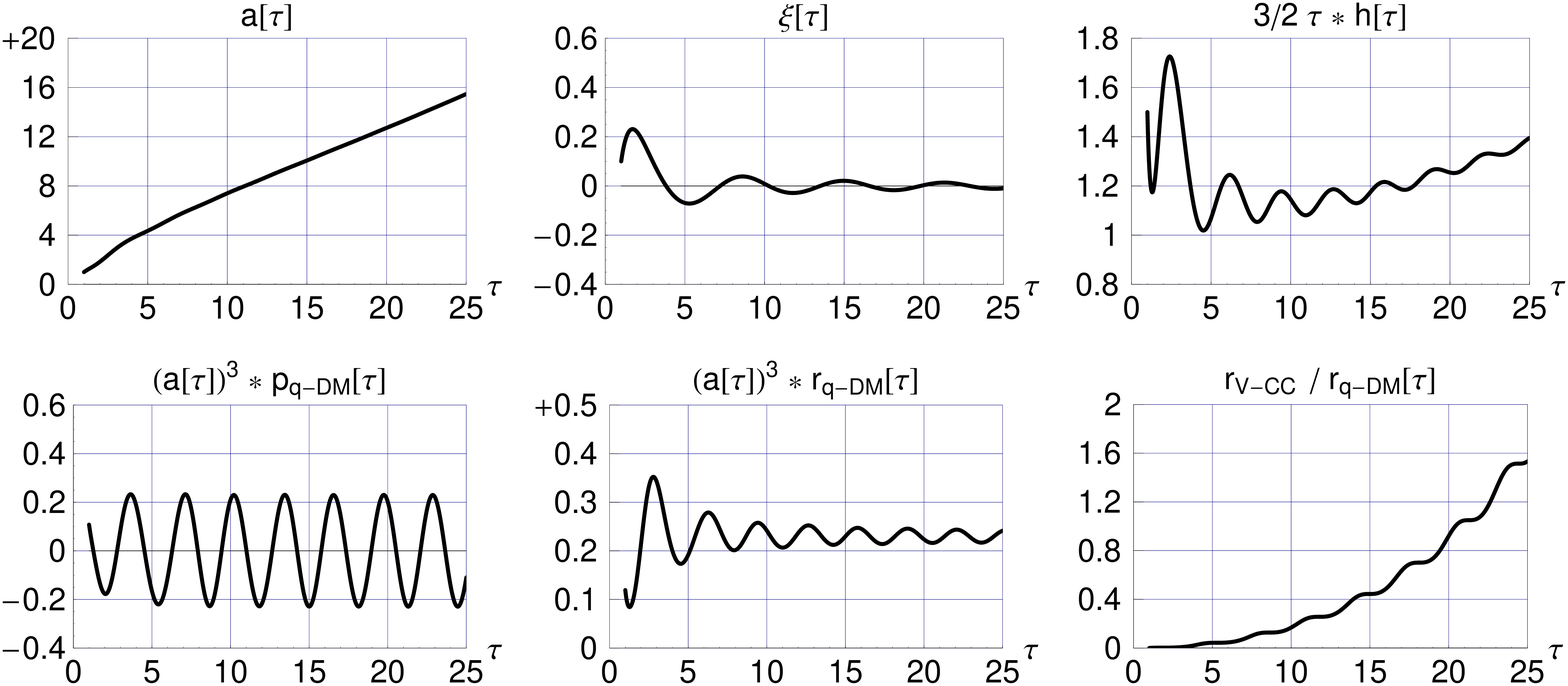}
\end{center}
\vspace*{-5mm}
\caption{
Same as Fig.~\ref{fig:solution-equil-mu} but now with
$u=-1/3-1/10000$.
The boundary conditions for $\{a(1),\, \dot{a}(1),\, \xi(1) \}$ are the same
as in Fig.~\ref{fig:solution-equil-mu}, but the derived value
of the $\xi$ derivative from \eqref{eq:dimensionless-ODEs-Friedmann}
is somewhat different, $\dot{\xi}(1)=0.476598$.
The corresponding effective cosmological constant
from \eqref{eq:overline-r-V-CC} is nonvanishing,
with $r_{V-\text{CC}}=-p_{V-\text{CC}}\approx 10^{-4}$ for
$\xi_{\text{const},\,1}$ $\approx$ $-0.00010002$.
}
\label{fig:solution-nonequil-mu}
\vspace*{20cm}
\end{figure}

Numerical results for the case of an equilibrium value of the
chemical potential (dimensional $\mu=\mu_{0}$ and
dimensionless $u=u_{0}$)
are given in Fig.~\ref{fig:solution-equil-mu}.
The asymptotic value  of this $\xi(\tau)$ solution is given by
\beq
\label{eq:xi-const-equilibrium-mu}
\xi_{\text{const},\,1}= 0\,.
\eeq
The numerical results of Fig.~\ref{fig:solution-equil-mu} show that%
\bsubeqs\label{eq:r-q-DM-p-q-DM-results}
\beqa
\label{eq:r-q-DM-result}
\rho_\text{$q$-DM} &\propto& 1/a^3\,,
\\[2mm]
\label{eq:p-q-DM-result}
\langle a^3\,P_\text{$q$-DM} \rangle &\sim& 0\,,
\eeqa
\esubeqs
where the bracket $\langle\ldots \rangle$
in \eqref{eq:p-q-DM-result} denotes a time average
over a time interval very much larger than the Planck-scale
oscillation period of $\xi(\tau)$,
which is given  by $2\pi$ in our units.
Note that $h(\tau)$ and $a(\tau)$ oscillate with the
double frequency ($\omega=2\,M$)
or the half period (given by $\pi$ in our units),
and precisely these oscillations
may lead to particle production~\cite{Kofman-etal1997}.
The behavior \eqref{eq:r-q-DM-p-q-DM-results} corresponds to
what is expected for a cold-dark-matter component.
For the case considered, the
effective cosmological constant from \eqref{eq:overline-r-V-CC}
vanishes, $r_{V-\text{CC}}=-p_{V-\text{CC}}=0$.

Numerical results for the case of a nonequilibrium value of the
chemical potential (dimensionless $u = u_{0} - 10^{-4}$)
are given in Fig.~\ref{fig:solution-nonequil-mu}.
The asymptotic value  of this $\xi(\tau)$ solution is given by
\beq
\label{eq:xi-const-nonequilibrium-mu}
\xi_{\text{const},\,1}
={\sqrt{2}}\,\cos \left( \frac13\,\arccos
 \left[\frac{-10003}{10000\,{\sqrt{2}}}\right]\right) -1
\approx -0.00010002\,,
\eeq
where the arccosine function for real arguments in $[-1,\,1]$
takes values in $[0,\,\pi]$.
The top-right panel for $(1.5\;\tau) \times h$
in Fig.~\ref{fig:solution-nonequil-mu} illustrates the
onset of de-Sitter-type expansion ($h=\text{const.}$)
for $\tau \gtrsim 20$. Indeed,
the effective cosmological constant from \eqref{eq:overline-r-V-CC}
is nonvanishing, $r_{V-\text{CC}}=-p_{V-\text{CC}}\approx 10^{-4}$,
and starts to dominate for $\tau \gtrsim 20$ 
(bottom right panel in Fig.~\ref{fig:solution-nonequil-mu}).
For $\tau \gg 20$,  the expansion is driven by the
effective cosmological constant, but the bottom-left and
bottom-middle panels of Fig.~\ref{fig:solution-nonequil-mu}
still show the behavior \eqref{eq:r-q-DM-p-q-DM-results}.

%%\newpage%%tmp
\section{Discussion}
\label{sec:Discussion}

The model universe of Fig.~\ref{fig:solution-nonequil-mu}
contains already the two main ingredients of our present
universe: a nonvanishing effective cosmological constant
with $\rho_{V-\text{CC}}=-P_{V-\text{CC}}>0$
and a cold-dark-matter component with the behavior
\eqref{eq:r-q-DM-p-q-DM-results}.
The behavior \eqref{eq:r-q-DM-p-q-DM-results} has been %%AAA
established numerically in Sec.~\ref{sec:Numeric-solutions}
but can also be shown analytically.

With the standard Einstein gravity from Eqs.~\eqref{eq:Einstein-eq}
and \eqref{eq:energy-momentum-tensor-q-followup},
the effective pressureless fluid from the rapidly-oscillating
part of the $q$-field can be expected to cluster
gravitationally in the same way as a hypothetical
cold-dark-matter particle would do
(see, e.g., Chap.~15 of Ref.~\cite{Peacock1999}).
Remaining issues are the addition of ``standard'' matter
(as mentioned in the first paragraph of Sec.~\ref{sec:Cosmological-model})
and the effects from particle production induced by the ultra-rapid
(and initially large-amplitude) oscillations
of the $q$-field~\cite{Kofman-etal1997}.

A more realistic description than Fig.~\ref{fig:solution-nonequil-mu}
requires a very much smaller (but nonzero) value of
$|\mu - \mu_{0}|/q_{0}$, so that the cross-over from
Friedmann--Robertson--Walker-type expansion
(Hubble parameter $H \equiv \dot{a}/a \propto 1/t$)
to de-Sitter-type expansion ($H = \text{constant}$)
occurs at a cosmic age of the order of $10^{9}$ years,
instead of $20 \times \hbar/E_{P} \sim 10^{-42}\;\text{s}$
as in Fig.~\ref{fig:solution-nonequil-mu}.

The main challenge is to \emph{derive} the correct
nonzero value of $|\mu - \mu_{0}|/q_{0}$.
One mechanism would be a kick of the vacuum energy density $\rho_{V}(q)$
by $\text{TeV}$-mass particle decays~\cite{KV2009-electroweak},
now in a theory with $G=\text{const}$ and $q$-derivatives.
But it is very well possible that another mechanism operates
to give a nonzero asymptotic value of $\rho_{V}(q)$,
that is, under the assumption that the $q$-theory approach
is relevant to dark energy and dark matter.

%%%%\newpage%%tmp
\section*{\hspace*{-5mm}ACKNOWLEDGMENTS}
\noindent
FRK thanks V. Emelyanov for an interesting discussion.
The work of GEV has been supported by the European Research Council
(ERC) under the European Union's Horizon 2020 research and innovation programme 
(Grant Agreement No. 694248).

\end{document}